\documentclass[prb,amsmath,amssymb]{revtex4}
\usepackage{graphicx}
\usepackage{bm}

\def\br{ \bm{r} }
\def\bR{ \bm{R} }
\def\bk{ \bm{k} }
\def\bq{ \bm{q} }
\def\bvs{ \bm{v}_s }
\def\im{\,\mathrm{Im}\,}

\def\Tr{\mathrm{Tr}}
\def\tr{\mathrm{tr}}
\def\Det{\mathrm{Det}\,}

\begin{document}

\title{Classical phase fluctuations in $d$-wave superconductors}

\author{K. V. Samokhin and B. Mitrovi\'c}

\address{Department of Physics, Brock University,
St.Catharines, Ontario, Canada L2S 3A1}
\date{\today}

\begin{abstract}
We study the effects of low-energy nodal quasiparticles on the
classical phase fluctuations in a two-dimensional $d$-wave
superconductor. The singularities of the phase-only action at
$T\to 0$ are removed in the presence of disorder, which justifies
using an extended classical $XY$-model to describe phase
fluctuations at low temperatures.
\end{abstract}

\maketitle

\section{Introduction}
\label{sec:Intro}

The spectacular successes of the Bardeen-Cooper-Schrieffer (BCS)
mean-field theory of superconductivity are based on the fact that
the so-called Ginzburg-Levanyuk number $Gi_{(D)}$, which controls
the size of fluctuation effects in $D$-dimensional samples, is
very small in bulk conventional materials.\cite{LarVar05} On the
other hand, the order parameter fluctuations become more
pronounced, even dominant, in low-dimensional systems with a small
Fermi energy, e.g. in quasi-two-dimensional cuprate
superconductors.\cite{KBCC88} In particular, the fluctuations of
the order parameter phase in the underdoped cuprates are enhanced
due to a low value of the superfluid density, leading to the large
deviations from the BCS picture, including the pseudogap
phenomenon.\cite{TS99} According to Emery and
Kivelson,\cite{EK95-2} the Cooper pairs survive in underdoped
cuprates far above the critical temperature $T_c$, but without
global phase coherence, which is destroyed by thermal phase
fluctuations through the Berezinskii-Kosterlitz-Thouless
mechanism.\cite{BKT} This idea has been further elaborated by many
authors, for a review see, e.g., Ref. \cite{LQS01}. As temperature
is lowered, a long-range phase coherence sets in, but the phase
fluctuations continue to play important role in the
superconducting state, remaining predominantly classical well
below $T_c$.\cite{EK95-1,CKEM99} One might expect however that
quantum phase fluctuations eventually take over at the lowest
temperatures. The crossover temperature $T_{cl}$ between the
classical and quantum regimes is quite high (of the order of
$T_c$) in clean charged systems, but can be significantly reduced
in the presence of dissipation. While the estimates of $T_{cl}$ in
cuprates obtained by different groups vary considerably, see e.g.
Refs. \cite{EK95-1,KDH01,BCCPR01}, here we adopt the view that the
dissipation is strong enough to make the quantum effects
negligible.

In this brief review we develop an effective long-wavelength
theory of the classical phase fluctuations in $d$-wave
superconductors, both with and without elastic disorder. At the
Gaussian level, the lowest-order term in the gradient energy
expansion is simply $\rho_s\bvs^2/2$, where $\rho_s$ is the
superfluid mass density and $\bvs$ is the superfluid velocity. We
show that the higher-order gradient terms, which contain
$\nabla\bvs$, are singular in a clean system at low temperatures
due to the presence of gap zeros,\cite{SM03}  and also discuss the
effects of disorder on those singularities. As a by-product of our
theory, we address the question whether using the classical
$XY$-model in $d$-wave superconductors can be justified from
microscopic theory.

The article is organized as follows. In Sec. \ref{sec:Derivation},
the general field-theoretical description of the bosonic
excitations in superconductors and a phase-only effective action
are derived in the clean case. In Sec. \ref{sec:2D}, we focus on
the case of a two-dimensional neutral $d$-wave superconductor,
which is treated in the nodal approximation. The microscopic
expressions for the energy of fluctuations are compared to the
predictions of the classical $XY$-model. In Sec.
\ref{sec:Disorder}, we derive the $d$-wave phase-only action in
the presence of elastic impurity scattering. Sec.
\ref{sec:Conclusion} concludes with a discussion of our results.

\section{Effective field theory: Clean case}
\label{sec:Derivation}

The starting point of our analysis is the tight-binding
Hamiltonian
\begin{equation}
\label{Hamilt}
    H=\sum\limits_{\br\br'}\xi_{\br\br'}c^\dagger_{\br\sigma}
    c_{\br'\sigma}+\sum\limits_{\br}U_{\br}c^\dagger_{\br\sigma}
    c_{\br\sigma}
    -g\sum\limits_{\langle\br\br'\rangle}B^\dagger_{\br\br'}B_{\br\br'}
    +\frac{1}{2}\sum\limits_{\br\br'}(n_{\br}-n_0)
    V_{\br\br'}(n_{\br'}-n_0),
\end{equation}
where $\br$ label the sites of a tetragonal lattice. The hopping
amplitude $t_{\br\br'}$ and the chemical potential $\mu$ are
combined into the band-dispersion matrix
$\xi_{\br\br'}=-t_{\br\br'}-\mu\delta_{\br\br'}$, which is real
and symmetric in the absence of external magnetic field. The
second term describes impurity scattering. The third term is the
BCS interaction in the $d$-wave channel, $g>0$ is the coupling
constant, and the operator $B_{\br\br'}=(c_{\br'\downarrow}
c_{\br\uparrow}-c_{\br'\uparrow}c_{\br\downarrow})/\sqrt{2}$
destroys a singlet pair of electrons at the nearest-neighbor sites
$\langle\br\br'\rangle$ in the $xy$ plane. The last term describes
the repulsive interaction between electrons,
$n_{\br}=c^\dagger_{\br\sigma}c_{\br\sigma}$ is the particle
number density, $n_0=\langle n_{\br}\rangle$ is the average number
of particles per site (which is equal to the ionic background
density, thus ensuring the overall charge neutrality of the
system), and $V_{\br\br'}$ is the interaction matrix.

Let us first look into the clean case. Setting $U_{\br}=0$ in Eq.
(\ref{Hamilt}), the partition function can be written as a
functional integral over Grassmann fields $c_{\br\sigma}(\tau)$
and $\bar c_{\br\sigma}(\tau)$:
\begin{equation}
\label{Zcc}
    Z=\Tr\, e^{-\beta H}=\int{\cal D}c\,{\cal D}\bar c\;e^{-S[\bar c,c]},
\end{equation}
where $S=\int_0^\beta d\tau\left[\sum_{\br}\bar
c_{\br\sigma}\partial_\tau c_{\br\sigma}+H(\tau)\right]$,
$\beta=1/T$ (in our units $k_B=\hbar=1$). Using the
Hubbard-Stratonovich transformation to decouple the interaction
terms, we end up with the representation of $Z$ as a functional
integral over $c,\bar c$ and two bosonic fields: a complex field
$\Delta_{\br\br'}(\tau)$, which describes the superconducting
order parameter fluctuations and is non-zero only on the bonds
between the nearest neighbors, and a real scalar potential field
$\varphi_{\br}(\tau)$. The fermionic part of the action then
becomes $S=\Tr(\bar C{\cal G}^{-1}C)$, where
$$
\label{Nambu C}
    C_{\br}=\left(%
    \begin{array}{c}
    c_{\br\uparrow} \\
    \bar c_{\br\downarrow} \\
    \end{array}\right),\quad
    \bar C_{\br}=\left(%
    \begin{array}{cc}
    \bar c_{\br\uparrow} & c_{\br\downarrow} \\
    \end{array}%
\right)
$$
are the Nambu spinors, and ${\cal G}^{-1}$ is the inverse Green's
operator:
\begin{equation}
\label{Gdef}
    {\cal G}^{-1}(\br,\tau;\br',\tau')
    =\delta(\tau-\tau')
    \left(\begin{array}{cc}
    \delta_{\br\br'}[-\partial_{\tau}-i\varphi_{\br}(\tau)]-\xi_{\br\br'} &
    -\Delta_{\br\br'}(\tau)\\
    -\Delta^*_{\br\br'}(\tau) &
    \delta_{\br\br'}[-\partial_{\tau}+i\varphi_{\br}(\tau)]
    +\xi_{\br\br'}
    \end{array}\right).
\end{equation}
We use the notation ``$\Tr$'' for the full operator trace with
respect to both the space-time coordinates and the Nambu matrix
indices, reserving ``$\tr$'' for a $2\times 2$ matrix trace in the
Nambu space. Integrating out the fermionic fields, we obtain
\begin{equation}
\label{Z}
    Z=\int{\cal D}\Delta^*\,{\cal D}\Delta\,{\cal D}\varphi\,
    \;e^{-S_{eff}[\Delta^*,\Delta,\varphi]},
\end{equation}
with the effective action
\begin{equation}
\label{Seff gen}
    S_{eff}=-\Tr\ln {\cal G}^{-1}
    +\int\limits_0^\beta d\tau\left(\frac{1}{g}
    \sum\limits_{\br\br'}|\Delta_{\br\br'}|^2
    +\frac{1}{2}\sum\limits_{\br\br'}\varphi_{\br}
    V_{\br\br'}^{-1}\varphi_{\br'}
    -in_0\sum\limits_{\br}\varphi_{\br}\right).
\end{equation}

The mean-field BCS theory corresponds to a stationary and uniform
saddle point of the effective action (\ref{Seff gen}), which is
found from the equations $\delta S_{eff}/\delta\Delta^*=\delta
S_{eff}/\delta\varphi=0$. The solution describing $d$-wave pairing
is given by $\varphi_{0,\br}=0$ and
\begin{equation}
\label{Delta dwave}
    \Delta_{0,\br\br'}=\left\{\begin{array}{cc}
    +\Delta_0, & \quad\mbox{if}\quad\br'=\br\pm a\hat{\bm{x}}, \\
    -\Delta_0, & \quad\mbox{if}\quad\br'=\br\pm a\hat{\bm{y}}.
    \end{array}\right.
\end{equation}
In the momentum representation,
$\Delta_{\bk}=\Delta_0(T)\phi_{\bk}$, where $\phi_{\bk}=2(\cos
k_xa-\cos k_ya)$ is the $d$-wave symmetry factor. The temperature
dependence of the gap amplitude $\Delta_0$ is determined by the
standard BCS self-consistency equation, generalized to the case of
an anisotropic order parameter.\cite{Book}

Inverting the operator (\ref{Gdef}), we find the mean-field matrix
Green's function:
\begin{equation}
\label{G0}
    {\cal G}_0(\bk,\omega_n)=
    \left(%
    \begin{array}{cc}
    G_0(\bk,\omega_n) & -F_0(\bk,\omega_n) \\
    -F_0(\bk,\omega_n) & -G_0(-\bk,-\omega_n) \\
    \end{array}%
    \right)=
    -\frac{i\omega_n\tau_0+\xi_{\bk}\tau_3+\Delta_{\bk}\tau_1}{\omega_n^2
    +\xi_{\bk}^2+\Delta_{\bk}^2}.
\end{equation}
Here $\omega_n=(2n+1)\pi T$ is the fermionic Matsubara frequency,
$G_0$ and $F_0$ are the usual normal and anomalous Gor'kov's
functions of the superconductor,\cite{AGD} $\xi_{\bk}=\xi_{-\bk}$
is the band dispersion of free electrons, and $\tau_i$ are Pauli
matrices in the Nambu space. The Green's function (\ref{G0})
determines the single-particle properties in the mean-field
approximation. In particular, after its analytical continuation to
the real frequency axis, $i\omega_n\to\omega+i0$, one obtains the
energies of elementary fermionic excitations, or the Bogoliubov
quasiparticles:
\begin{equation}
\label{E_k}
    E_{\bk}=\sqrt{\xi_{\bk}^2+\Delta_{\bk}^2}.
\end{equation}
For the $d$-wave order parameter, the gap in the excitation energy
vanishes along the diagonals of the Brillouin zone. These zeros,
or the gap ``nodes'', are responsible for many peculiar
thermodynamical and transport properties in the superconducting
state on the mean-field level.\cite{Book} Below we show that the
gap nodes also have dramatic effects on the long-wavelength
behavior of phase fluctuations.

\subsection{Fluctuations}
\label{sec:Fluct}

Deviations from the mean-field solution can be represented in
terms of the amplitude and phase fluctuations:
$\Delta_{\br\br'}(\tau)=[\Delta_{0,\br\br'}+\delta\Delta_{\br\br'}(\tau)]
e^{i\Theta_{\br\br'}(\tau)}$, where $\delta\Delta$ and $\Theta$
are real. We neglect the amplitude fluctuations because they are
gapped, see e.g. Ref. \cite{BTCC02}, and therefore make a
negligible contribution at low temperatures. Since the number of
bonds in a square lattice is twice the number of sites, one needs
two on-site phase fields $\theta_{\br}(\tau)$ and
$\tilde\theta_{\br}(\tau)$ to describe the phase degrees of
freedom. One possible parametrization is
\begin{equation}
    \Theta_{\br\br'}=\left\{\begin{array}{cc}
      \theta_{\br}, & \mbox{if }\br'=\br+a\hat{\bm{x}} \\
      \theta_{\br}+\tilde\theta_{\br}, & \mbox{if }
      \br'=\br+a\hat{\bm{y}}.
    \end{array}\right.
\end{equation}
The fluctuations of $\tilde\theta$, which describe a change in the
symmetry of the order parameter from a pure $d$-wave to a
$d+is$-wave, can be neglected.\cite{PRRM00}

If $\theta_{\br}$ changes slowly over the lattice constant, then
one can make the replacement
\begin{equation}
\label{average phase}
    \Delta_{0,\br\br'}e^{i\theta_{\br}}\to\Delta_{0,\br\br'}
    e^{i(\theta_{\br}+\theta_{\br'})/2},
\end{equation}
(recall that $\br$ and $\br'$ are nearest neighbors). The next
step is to perform a gauge transformation to make the off-diagonal
elements of ${\cal G}^{-1}$ in Eq. (\ref{Gdef}) real:
\begin{equation}
\label{tilde G}
    U^\dagger(\br,\tau){\cal G}^{-1}(\br,\tau;\br',\tau')U(\br',\tau')=
     \tilde {\cal G}^{-1}(\br,\tau;\br',\tau'),
\end{equation}
where $U(\br,\tau)=\exp[i\tau_3\theta_{\br}(\tau)/2]$. This
transformation leaves the operator trace in the effective action
(\ref{Seff gen}) invariant. Although the order parameter
(\ref{average phase}) is no longer invariant under local phase
rotations $\theta_{\br}\to\theta_{\br}+2\pi$, our results are not
affected since we consider only small fluctuations of the phase in
the low-temperature limit, where the contribution of vortices can
be safely neglected.

The operator $\tilde{\cal G}^{-1}$ can be represented in the form
$\tilde {\cal G}^{-1}={\cal G}_0^{-1}-\Sigma$, where ${\cal G}_0$
is the mean-field matrix Green's function, whose Fourier transform
is given by Eq. (\ref{G0}), and
\begin{equation}
\label{Sigma}
    \Sigma(\br,\tau;\br',\tau')
    =\delta(\tau-\tau')\tau_3\biggl\{i\delta_{\br\br'}\left[\frac{1}{2}
    \frac{\partial\theta_{\br}(\tau)}{\partial\tau}+\varphi_{\br}(\tau)
    \right]
    +\xi_{\br\br'}\left[e^{-i\tau_3[\theta_{\br}(\tau)-\theta_{\br'}(\tau)]/2}
    -1\right]\biggr\}
\end{equation}
is the self-energy correction due to fluctuations. At slow
temporal and spatial variations of $\theta$ and small $\varphi$,
one can expand the effective action (\ref{Seff gen}) in powers of
$\Sigma$, keeping only the two lowest orders in the expansion,
with the following result:
\begin{equation}
\label{Seff theta phi}
    S_{eff}[\theta,\varphi]=S_0
    +\Tr({\cal G}_0\Sigma)+\frac{1}{2}\Tr({\cal G}_0\Sigma{\cal
    G}_0\Sigma)+O(\Sigma^3)
    +\frac{1}{2}\int\limits_0^\beta d\tau
    \sum\limits_{\br\br'}\varphi_{\br}(\tau)
    V_{\br\br'}^{-1}\varphi_{\br'}(\tau)-in_0\int\limits_0^\beta
    d\tau\sum\limits_{\br}\varphi_{\br}(\tau),
\end{equation}
where
\begin{equation}
    S_0=-\Tr\ln{\cal G}_0^{-1}+\beta\frac{1}{g}
    \sum\limits_{\br\br'}|\Delta_{0,\br\br'}|^2=\beta{\cal E}_0
\end{equation}
is the saddle-point action, ${\cal E}_0$ is the total mean-field
energy of the superconductor, and ${\cal G}_0$ is the saddle-point
Green's function, see Eq. (\ref{G0}). For non-interacting
fluctuations we keep only the terms of the first and second order
in $\Sigma$.

Calculating the traces in Eq. (\ref{Seff theta phi}), we obtain
the Gaussian action
\begin{equation}
    S_{eff}[\theta,\varphi]=S_0+S_1+S_2,
\end{equation}
where
\begin{equation}
    S_1=\frac{in_0}{2}\int\limits_0^\beta d\tau\;\sum\limits_{\br}
    \frac{\partial\theta_{\br}(\tau)}{\partial\tau}=
    i\pi
    n_0\sum\limits_{\br}\frac{\theta_{\br}(\beta)-\theta_{\br}(0)}{2\pi}
\end{equation}
is the topological term containing the phase winding numbers, and
\begin{equation}
\label{S2}
    S_2 = \frac{1}{2}\sum_Q\left[L_{\varphi\varphi}(Q)|\varphi(Q)|^2
    +L_{\theta\varphi}(Q)\theta^*(Q)\varphi(Q)
    +\frac{1}{4}L_{\theta\theta}(Q)|\theta(Q)|^2\right].
\end{equation}
Here we use the shorthand notations $Q=(\bq,\nu_m)$ and
$$
    \sum_Q (...)=T\sum\limits_m\sum\limits_{\bq} (...),
$$
where $\nu_m=2m\pi T$ is the bosonic Matsubara frequency and the
momentum summation goes over the first Brillouin zone. The lattice
Fourier transforms of the fields are defined by the usual
expressions: $\theta_{\br}(\tau)={\cal
N}^{-1/2}\sum_{\bq}\theta(\bq,\tau)e^{i\bq\tau}$ \emph{etc}, where
${\cal N}$ is the number of lattice sites. Since both $\varphi$
and $\theta$ are real, they satisfy $\theta^*(Q)=\theta(-Q)$,
$\varphi^*(Q)=\varphi(-Q)$. The coefficients in Eq. (\ref{S2}) are
given by
\begin{eqnarray}
    && L_{\varphi\varphi}(Q)=V^{-1}(\bq)+\Pi_0(Q),\nonumber \\
    &&\label{Ls} L_{\theta\varphi}(Q)=i\nu_m\Pi_0(Q)+q_i\Pi^i_1(Q),\\
    && L_{\theta\theta}(Q)=\nu_m^2\Pi_0(Q)-
    2i\nu_mq_i\Pi^i_1(Q)+q_i
    q_j\Pi^{ij}_2(Q).\nonumber
\end{eqnarray}
Here $V(\bq)$ is the Fourier transform of the interaction matrix
$V_{\br\br'}$, and
\begin{eqnarray}
  &&\label{Pi0}\Pi_0(Q)=-\sum_K\tr[{\cal G}_0(K+Q)\tau_3{\cal G}_0(K)\tau_3],\\
  &&\label{Pi1}
    \Pi^i_1(Q)=\sum_K v_i\,\tr[{\cal G}_0(K+Q)\tau_3{\cal G}_0(K)\tau_0],\\
  &&\label{Pi2}\Pi^{ij}_2(Q)=\sum_K m^{-1}_{ij}\tr[{\cal G}_0(K)\tau_3]+
    \sum_K v_iv_j\,\tr[{\cal G}_0(K+Q)\tau_0{\cal G}_0(K)\tau_0].
\end{eqnarray}
In these expressions, $K=(\bk,\omega_n)$,
$$
    \sum_K (...)=T\sum\limits_n\frac{1}{\cal N}\sum\limits_{\bk}(...)
    \stackrel{{\cal N}\to\infty}{\longrightarrow}
    T\sum\limits_n\Omega\int\frac{d^D\bk}{(2\pi)^D} (...),
$$
$m^{-1}_{ij}(\bk)=\partial^2\xi_{\bk}/\partial k_i\partial k_j$ is
the inverse effective mass tensor,
$\bm{v}(\bk)=\partial\xi_{\bk}/\partial\bk$ is the quasiparticle
band velocity, and $\Omega$ is the unit cell volume (to simplify
the notations, below we set $\Omega=1$). Eqs. (\ref{Ls}) are
obtained in the limit of small $\bq$ from more general
expressions, using the gradient expansion
$\xi_{\bk+\bq}=\xi_{\bk}+\bm{v}\bq+(1/2)m^{-1}_{ij}q_i
q_j+O(q^3)$. Integrating out the field $\varphi$, we finally
arrive at the phase-only effective action:
\begin{equation}
\label{Seff theta}
    S_{eff}[\theta]=S_0+S_1+\frac{1}{8}\sum_Q
    \biggl[L_{\theta\theta}(Q)+
    \frac{L^2_{\theta\varphi}(Q)}{L_{\varphi\varphi}(Q)}\biggr]|\theta(Q)|^2.
\end{equation}

In this article, we focus on the case of classical phase
fluctuations and neglect all interactions other than those
responsible for the Cooper pairing, which corresponds to
neglecting the $\tau$-dependence of $\theta$ and setting
$V(\bq)=0$. Then the topological term vanishes and only the
$\Pi_2^{ij}$ contribution survives in the third term in Eq.
(\ref{Seff theta}), so that the effective action becomes
\begin{equation}
\label{Seff cl}
    S_{eff}[\theta]=\beta{\cal E}_0+\beta{\cal E}[\theta],
\end{equation}
where ${\cal E}$ is the energy of fluctuations in the Gaussian
approximation. Calculating the Matsubara sums in Eq. (\ref{Pi2})
and introducing the superfluid velocity $\bvs=(1/2m)\nabla\theta$,
where $m$ is the electron mass, we obtain
\begin{equation}
\label{E gen}
    {\cal E}=\frac{1}{2}\sum\limits_{\bq}{\cal K}_{ij}(\bq)
    v^*_{s,i}(\bq)v_{s,j}(\bq),
\end{equation}
with the kernel
\begin{equation}
\label{kernel}
    {\cal K}_{ij}(\bq)\equiv m^2\Pi^{ij}_2(\bq,0)
    ={\cal K}^{(0)}_{ij}+{\cal I}_{ij}(\bq),
\end{equation}
where
\begin{eqnarray}
    &&\label{cal K0} {\cal K}^{(0)}_{ij}=
    2m^2T\sum\limits_n\int\frac{d^D\bk}{(2\pi)^D}\,
    m^{-1}_{ij}(\bk)G_0(\bk,\omega_n),\\
    &&\label{cal I} {\cal I}_{ij}(\bq)=-m^2\int\frac{d^D\bk}{(2\pi)^D}
    \,v_i(\bk)v_j(\bk)\biggl[C_-(\bk,\bq)\frac{\tanh\frac{E_{\bk+\bq}}{2T}+
    \tanh\frac{E_{\bk}}{2T}}{E_{\bk+\bq}+E_{\bk}}
    +C_+(\bk,\bq)\frac{\tanh\frac{E_{\bk+\bq}}{2T}-
    \tanh\frac{E_{\bk}}{2T}}{E_{\bk+\bq}-E_{\bk}}\biggr],
\end{eqnarray}
and
\begin{equation}
\label{Cpm}
    C_\pm(\bk,\bq)=\frac{1}{2}\left(1\pm
    \frac{\xi_{\bk}\xi_{\bk+\bq}+\Delta_{\bk}\Delta_{\bk+\bq}}{E_{\bk}E_{\bk+\bq}}\right)
\end{equation}
are the coherence factors.

An important characteristic of the superconductor is the
superfluid density tensor, which is defined as
\begin{equation}
    \rho_{s,ij}(T)\equiv{\cal K}_{ij}(\bm{0}).
\end{equation}
Its temperature dependence can be easily found in two limiting
cases. In the normal state, $\Delta_{\bk}=0$, and one can use the
identity $\partial G_0/\partial\bk=\bm{v}G_0^2$ in Eq. (\ref{Pi2})
to obtain $\rho_{s,ij}(T>T_c)=0$. On the other hand, at zero
temperature we have $\rho_{s,ij}(0)={\cal K}^{(0)}_{ij}$, since
${\cal I}_{ij}(\bm{0})=0$ at $T=0$.

The expressions (\ref{kernel},\ref{cal K0},\ref{cal I}) are valid
for arbitrary band structure and gap symmetry. In a
Galilean-invariant system, i.e. for $\xi_{\bk}=\bk^2/2m-\mu$, the
tensor (\ref{cal K0}) takes a particularly simple form: ${\cal
K}^{(0)}_{ij}=\rho_0\delta_{ij}$, where
$\rho_0=2mT\sum_n\int_{\bk}G_0(\bk,\omega_n)$ is the mass density
of electrons. Therefore, the superfluid density tensor at zero
temperature is $\rho_{s,ij}(0)=\rho_0\delta_{ij}$, i.e. all
electrons are superconducting. In general, there is no such simple
relation in a crystal.

\section{Two-dimensional case}
\label{sec:2D}

In this section, we apply the general theory developed above to a
two-dimensional $d$-wave superconductor. In this case, the
low-energy physics at $T\to 0$ can be conveniently described using
the so-called ``nodal approximation'',\cite{Lee93} which takes
advantage of the fact that the excitation energy (\ref{E_k}) for
the $d$-wave order parameter can be linearized in the vicinity of
the four gap nodes located at $\bk_n=k_F\hat\bk_n$ ($n=1,2,3,4$)
on the Fermi surface. Here
$$
    \displaystyle\hat\bk_1=\frac{\hat{\bm{x}}+\hat{\bm{y}}}{\sqrt{2}},\
    \displaystyle\hat\bk_2=\frac{-\hat{\bm{x}}+\hat{\bm{y}}}{\sqrt{2}},\
    \displaystyle\hat\bk_3=\frac{-\hat{\bm{x}}-\hat{\bm{y}}}{\sqrt{2}},\
    \displaystyle\hat\bk_4=\frac{\hat{\bm{x}}-\hat{\bm{y}}}{\sqrt{2}}.
$$
For the nodal quasiparticles in the vicinity of the $n$th node we
have $\bk=\bk_n+\delta\bk$, and
\begin{equation}
\label{nodal_approx}
    \xi_{\bk}=v_F\delta k_\perp,\quad
    \Delta_{\bk}=v_\Delta\delta k_\parallel,\quad
    E_{\bk}=\sqrt{v_F^2\delta k_\perp^2+v_\Delta^2
    \delta k_\parallel^2},
\end{equation}
where $\delta k_\perp$ and $\delta k_\parallel$ are the momentum
components perpendicular and parallel to the Fermi surface, $v_F$
is the Fermi velocity at the nodes, and
$v_\Delta=|\partial\Delta_{\bk}/\partial\bk_\parallel|$ is the
slope of the superconducting gap function near the nodes. Thus,
the excitation spectrum near the gap nodes is described by an
anisotropic Dirac cone. The anisotropy ratio $v_F/v_\Delta$ is an
important characteristic of the high-$T_c$ cuprates, which depends
on the material and the doping level, e.g. $v_F/v_\Delta\simeq 14$
and $19$ in the optimally doped YBCO and Bi-2212,
respectively.\cite{ratio}

In the nodal approximation, the momentum integration over the
whole Brillouin zone is replaced by a sum of four integrals over
the small regions in $\bk$-space around the nodes:
\begin{equation}
\label{nodal int}
    \int\frac{d^2\bk}{(2\pi)^2}\,(...)\to
    \sum\limits_{n=1}^4\int\frac{d\delta k_\perp d\delta
    k_\parallel}{(2\pi)^2}\,(...)
    =\frac{1}{2\pi v_Fv_\Delta}\sum\limits_{n=1}^4
    \int\limits_0^{\epsilon_{max}}d\epsilon\,
    \epsilon\int\limits_0^{2\pi}\frac{d\alpha}{2\pi}\,(...).
\end{equation}
In the last integral, we changed to the polar coordinates:
$v_F\delta k_\perp=\epsilon\cos\alpha$, $v_\Delta\delta
k_\parallel=\epsilon\sin\alpha$, and $E_{\bk}=\epsilon$. The
ultraviolet cutoff $\epsilon_{max}\simeq\Delta_0$ is introduced to
make sure that the area of the integration region is equal to the
area of the original Brillouin zone.\cite{DL00} In most
calculations in this article this cutoff can be extended to
infinity.

\subsection{Classical phase fluctuations}
\label{sec:2D-class}

We can now calculate the energy of the classical phase
fluctuations, see Eq. (\ref{E gen}). The nodal approximation
cannot be applied to the momentum integral in Eq. (\ref{cal K0})
because it contains contributions from all electrons, including
those far from the Fermi surface. Assuming that the band
dispersion can be treated in the effective mass approximation,
which amounts to the replacement
$m^{-1}_{ij}\to(1/m^*)\delta_{ij}$, one obtains
\begin{equation}
\label{cal K0 eff mass}
    {\cal K}^{(0)}_{ij}=\frac{m}{m^*}\rho_0\delta_{ij},
\end{equation}
where $\rho_0$ is the average mass density of electrons.

In contrast, the second term in the kernel ${\cal K}$ can be
calculated in the nodal approximation. Using Eqs.
(\ref{nodal_approx},\ref{nodal int}), we have
\begin{equation}
\label{cal I sum}
    {\cal I}_{ij}(\bq)=-\frac{m^2}{2\pi}\frac{v_F}{v_\Delta}
    \sum\limits_{n=1}^4\hat k_{n,i}\hat k_{n,j}S_n(\bq),
\end{equation}
where
$$
    S_1(\bq)=S_3(\bq)=Ts\left(\frac{\gamma_{1}}{T}\right),\quad
    S_2(\bq)=S_4(\bq)=Ts\left(\frac{\gamma_{2}}{T}\right).
$$
Here
\begin{equation}
\label{gamma12}
    \gamma_{1,2}(\bq)=\frac{1}{\sqrt{2}}\sqrt{v_F^2(q_x\pm q_y)^2+
    v_\Delta^2(q_x\mp q_y)^2}
\end{equation}
are the energies of the nodal quasiparticles with $\delta\bk=\bq$,
and the scaling function $s(x)$ is defined by an integral:
\begin{equation}
\label{s}
    s(x)=\int\limits_0^\infty dy\,y
    \int\limits_0^{2\pi}\frac{d\alpha}{2\pi}\,
    [f_+(x,y,\alpha)+f_-(x,y,\alpha)],
\end{equation}
where
$$
   f_\pm=\frac{1}{2}\left(1\pm\frac{y+
   x\cos\alpha}{\sqrt{x^2+y^2+2xy\cos\alpha}}\right)\frac{\tanh(\sqrt{x^2+y^2+2xy\cos\alpha}/2)\mp
   \tanh(y/2)}{\sqrt{x^2+y^2+2xy\cos\alpha}\mp y}.
$$
Note that the cutoff energy $\epsilon_{max}$ has been replaced by
infinity, due to the rapid convergence of the integrals. One can
show that the function $s(x)$ has the following asymptotics:
\begin{equation}
\label{asymp}
    s(x)=\left\{\begin{array}{ll}
    \displaystyle 2\ln 2+\frac{x^2}{24} &,\ \mbox{at }x\to 0\\
    \displaystyle \frac{\pi x}{8} &,\ \mbox{at }x\to\infty.
    \end{array}\right.
\end{equation}

After the summation over the four nodes in Eq. (\ref{cal I sum}),
one finally obtains
\begin{equation}
\label{K final}
    {\cal K}_{ij}(\bq)=\frac{m}{m^*}\rho_0\delta_{ij}-
    \frac{m^2}{2\pi}\frac{v_F}{v_\Delta}T\left(\begin{array}{cc}
    F_+(\bq) & F_-(\bq)\\
    F_-(\bq) & F_+(\bq)
    \end{array}\right)_{ij},
\end{equation}
where
$$
    F_\pm(\bq)=s\left[\frac{\gamma_{1}(\bq)}{T}\right]\pm
    s\left[\frac{\gamma_{2}(\bq)}{T}\right].
$$
The expression (\ref{K final}) is exact in the nodal
approximation, i.e. for the conical quasiparticle spectrum. In
terms of $\bq$, the applicability region of the nodal
approximation is $\gamma_{1,2}(\bq)\ll\Delta_0$. At higher
energies of quasiparticles, the deviations of the spectrum from
the linearized form (\ref{nodal_approx}) should be taken into
account.

We would like to note that in the nodal approximation,
$\Pi_1^i(\bq,0)=0$ and therefore $L_{\theta\varphi}(\bq,0)=0$, see
Eqs. (\ref{Ls}), (\ref{Pi1}). This means that the classical
fluctuation energy in two dimensions has the form (\ref{E gen})
with the kernel (\ref{K final}), even if the Coulomb interaction
is taken into account.

At $T=0$, using the large-$x$ asymptotics of $s(x)$ in Eq.
(\ref{asymp}), the kernel takes the form
\begin{equation}
\label{K zero T}
    {\cal K}_{ij}(\bq)=\frac{m}{m^*}\rho_0\delta_{ij}-
    \frac{m^2}{16}\frac{v_F}{v_\Delta}
    \left(\begin{array}{ccc}
    \gamma_1(\bq)+\gamma_2(\bq) &\ & \gamma_1(\bq)-\gamma_2(\bq)\\
    \gamma_1(\bq)-\gamma_2(\bq) &\ & \gamma_1(\bq)+\gamma_2(\bq)
    \end{array}\right)_{ij}.
\end{equation}
Setting $\bq=0$ here, we find the superfluid density:
$\rho_s(0)=(m/m^*)\rho_0$. We also see that the kernel is a
non-analytical function of $\bq$, which means that no gradient
expansion of the energy exists.

At finite temperatures and small $\bq$, such that
$\gamma_{1,2}(\bq)\ll T$, the small-$x$ asymptotics of $s(x)$
yields
\begin{equation}
\label{K nonzero T}
    {\cal K}_{ij}(\bq)=\left(\frac{m}{m^*}\rho_0
    -\frac{2\ln 2}{\pi}\frac{v_F}{v_\Delta}m^2T\right)\delta_{ij}
    -\frac{m^2}{48\pi}\frac{v_F}{v_\Delta}\frac{1}{T}
    \left(\begin{array}{ccc}
    (v_F^2+v_\Delta^2)(q_x^2+q_y^2) && 2(v_F^2-v_\Delta^2)q_xq_y\\
    2(v_F^2-v_\Delta^2)q_xq_y && (v_F^2+v_\Delta^2)(q_x^2+q_y^2)
    \end{array}\right)_{ij}.
\end{equation}
The first term describes the depletion of the superfluid density
due to the thermal excitation of quasiparticles:
\begin{equation}
\label{sf density T}
    \rho_s(T)=\rho_s(0)-\frac{2\ln 2}{\pi}\frac{v_F}{v_\Delta}m^2T,
\end{equation}
see also Refs. \cite{Lee93,LW97}, which explains a linear in $T$
increase of the magnetic penetration depth $\lambda(T)$ at low
temperatures,\cite{Gross86,AGR91} observed in high-$T_c$
cuprates.\cite{Hardy93}

The quadratic $\bq$-dependence of the expression (\ref{K nonzero
T}) implies that the kernel ${\cal K}_{ij}(\bR)$ in real space is
proportional to $\exp(-|\bR|/\tilde\xi)$, with the length
$\tilde\xi$ given by
\begin{equation}
\label{tilde xi}
    \tilde\xi(T)=\left(\frac{m^2v_F^3}{48\pi
    v_\Delta\rho_{s}T}\right)^{1/2}\sim\left(\frac{T_c}{T}
    \right)^{1/2}\xi_0,
\end{equation}
where $\xi_0=v_F/2\pi T_c$ is the BCS coherence length (we assumed
that $v_F\gg v_\Delta$). This behavior is similar to that of the
electromagnetic response function in conventional $s$-wave
superconductors, see e.g. Ref. \cite{Tink96}. An important
difference however is that the characteristic length $\tilde\xi$
is temperature-dependent: $\tilde\xi(T)\sim T^{-1/2}$ at $T\to 0$.
It is because of the divergence of $\tilde\xi$ that the gradient
expansion of the classical energy of fluctuations breaks down at
$T\to 0$. Note also that the length $\tilde\xi$ is different from
other characteristic lengths discussed in the literature: $\xi_0$
-- the coherence length, or the correlation length of the gap
amplitude fluctuations, which remains constant at $T\to 0$, and
$\xi_{pair}$ -- the size of a Cooper pair, which is infinite in
the $d$-wave case.\cite{BTCC02} In a conventional $s$-wave
superconductor, all three lengths are of the same order.

The physical interpretation of our findings is the same as that of
the non-local Meissner effect:\cite{non-loc} since the gap
function $\Delta_{\bk}$ has nodes on the Fermi surface, then the
anisotropic coherence length $v_F/|\Delta_{\bk}|$ exceeds the
London penetration depth $\lambda_L$ close to the nodes, and the
local electrodynamics breaks down.

One can expect that the nodal quasiparticles also affect the
non-Gaussian terms in the effective action (\ref{Seff theta phi}).
As shown in the Appendix, indeed no expansion of the classical
fluctuation energy in powers of $\bvs$ exists at $T=0$.

\subsection{Failure of the classical $XY$-model}
\label{sec:2D-XY}

The effects of the phase fluctuations in superconductors are most
often studied using either the classical or the quantum versions
of the $XY$-model. The energy of the classical $XY$-model in the
absence of external fields has the form
\begin{equation}
\label{XYdef}
    {\cal E}_{XY}=\sum\limits_{\bR\bR'}J_{\bR\bR'}
    [1-\cos(\theta_{\bR}-\theta_{\bR'})],
\end{equation}
where $\theta_{\bR}$ is the phase of the order parameter at site
$\bR$ of a coarse-grained square lattice, whose lattice spacing
$d$ is of the order of the superconducting correlation length
$\xi$. The summation goes over all bonds in the lattice, and the
coupling constants $J_{\bR\bR'}=J(\bm{\rho})$, where
$\bm{\rho}=\bR-\bR'$, are called the phase stiffness coefficients.
While in the Gaussian approximation, see below, the coupling
constants are temperature-independent, the interaction of
fluctuations leads to a thermal renormalization of the $J$s, and
eventually to a phase transition into the disordered state.

The $XY$-model is believed to provide a correct description of any
system with broken U(1) symmetry if the amplitude fluctuations of
the order parameter are negligible. Typical examples are classical
Heisenberg magnets, superfluids and superconductors. The
experimental systems to which the lattice model (\ref{XYdef}) has
been applied include granular superconductors and fabricated
arrays of Josephson junctions, see, e.g., Ref. \cite{GW84} and the
references therein. In those cases $d$ is given by the distance
between grains. Although the simplicity of the $XY$-model is
physically appealing, its rigorous microscopic derivation for
homogeneous high-$T_c$ superconductors does not exist. The usual
way of justification, see, e.g., Ref. \cite{PRRM00}, involves
expanding the cosine in Eq. (\ref{XYdef}) and matching the
expansion coefficients with those in the Gaussian phase-only
action. That the microscopic theory fails to reproduce the quantum
generalization of the $XY$-model has already been noticed in Ref.
\cite{BTC04}. Here we show, following Ref. \cite{SM03}, that even
for the classical phase fluctuations in a $d$-wave superconductor
at low temperatures, the long-wavelength limit of the microscopic
theory is not consistent with Eq. (\ref{XYdef}).

For slow variations of the phase, the energy (\ref{XYdef}) takes a
Gaussian form:
$$
    {\cal E}_{XY}=\frac{1}{2}\sum\limits_{\bR\bR'}J_{\bR\bR'}
    (\theta_{\bR}-\theta_{\bR'})^2=\frac{1}{2}\sum\limits_{\bq}|\theta(\bq)|^2
    \sum\limits_{\bm{\rho}}J(\bm{\rho})(1-\cos\bq\bm{\rho}).
$$
In terms of the superfluid velocity, we have
\begin{equation}
\label{E XY Gauss}
    {\cal E}_{XY}=\frac{1}{2}\sum\limits_{\bq}{\cal K}^{XY}_{ij}(\bq)
    v^*_{s,i}(\bq)v_{s,j}(\bq).
\end{equation}
The kernel here has a well-defined Taylor expansion in powers of
$\bq$:
\begin{equation}
\label{XY kernel}
    {\cal K}^{XY}_{ij}(\bq)=\rho^{XY}_{s,ij}+\Lambda_{ij,kl}q_kq_l+O(\bq^4),
\end{equation}
where
\begin{equation}
    \rho^{XY}_{s,ij}=2m^2\sum\limits_{\bm{\rho}}J(\bm{\rho})\rho_i\rho_j
\end{equation}
is the superfluid mass density tensor (for example, if the only
non-zero coupling is between the nearest-neighbor sites, then
$\rho^{XY}_{s,ij}=8m^2d^2J\delta_{ij}$), and
\begin{equation}
\label{Lambda}
    \Lambda_{ij,kl}=-\frac{m^2}{6}
    \sum\limits_{\bm{\rho}}J(\bm{\rho})\rho_i\rho_j\rho_k\rho_l.
\end{equation}

Comparing Eqs. (\ref{XY kernel}) and (\ref{K zero T}), we see that
the effective long-wavelength theory of the classical phase
fluctuations at $T=0$ does not have the form of the $XY$-model,
since the momentum dependence of the two energies is clearly
different. At $T>0$, although the expression (\ref{K nonzero T})
is quadratic in $\bq$, the coefficients diverge as $T\to 0$, which
is not the case for $\Lambda_{ij,kl}$ above. Thus, the microscopic
theory fails to reproduce the long-wavelength structure of the
classical $XY$-model in a clean $d$-wave superconductor.

\section{Disordered case}
\label{sec:Disorder}

In the presence of impurities, a full effective field theory for
the disordered interacting system described by the Hamiltonian
(\ref{Hamilt}) would include the fluctuations of the order
parameter and of the scalar potential coupled with the
disorder-induced soft modes (the diffusons and the Cooperons).
Such theories, usually having the form of a non-linear
$\sigma$-model, have been developed, see, e.g. Ref.
\cite{sigma-models} and the references therein, to study the
effects that are beyond the scope of the present work, for
instance the suppression of $T_c$ due to the interplay of disorder
and interactions in $s$-wave superconductors. Our goal here is to
check if the elastic impurity scattering removes the divergencies
in the gradient expansion of the classical phase-only action
discussed above. The disorder is treated essentially in the
saddle-point approximation and the Coulomb interaction is
neglected.

As a bookkeeping device to obtain an effective action for the
order parameter fluctuations, we use the replica trick:
$\langle\ln Z\rangle=\lim_{n\to 0}(\langle Z^n\rangle-1)/n$ (the
angular brackets denote averaging with respect to disorder). From
Eq. (\ref{Z}) we have
\begin{equation}
\label{Zcc n}
    Z^n=\int\prod\limits_{a=1}^n{\cal D}c^a{\cal D}\bar c^a\, e^{-S[\bar c,c]},
\end{equation}
where $S=S_0+S_{int}$,
\begin{eqnarray}
\label{tilde S}
    &&S_0=\sum\limits_a\int\limits_0^\beta d\tau\biggl(\sum_{\br}\bar
    c^a_{\br\sigma}\partial_\tau c^a_{\br\sigma}+
    \sum\limits_{\br\br'}\xi_{\br\br'}\bar c^a_{\br\sigma}c^a_{\br'\sigma}
    +\sum\limits_{\br}U_{\br}\bar
    c^a_{\br\sigma}c^a_{\br\sigma}\biggr),\\
    &&S_{int}=-g\sum\limits_a\int\limits_0^\beta d\tau
    \sum\limits_{\langle\br\br'\rangle}\bar B^a_{\br\br'}B^a_{\br\br'}.
\end{eqnarray}
The impurity potential here is assumed to be Gaussian-distributed,
with zero mean and the correlator
$$
    \langle U_{\br}U_{\br'}\rangle=\frac{1}{2\pi
    N_F\tau}\delta_{\br\br'},
$$
where $N_F$ is the density of states at the Fermi level, and
$\tau$ is the electron mean-free time due to elastic scattering.
The next step is to use an incomplete Hubbard-Stratonovich
transformation to decouple only the interaction terms $S_{int}$ in
each replica, before disorder averaging.\cite{BK94} Proceeding as
in Sec. \ref{sec:Derivation}, we have
\begin{equation}
\label{Zn}
    \langle Z^n\rangle=\int\prod\limits_{a=1}^n{\cal D}\Delta^{a,*}\,{\cal D}\Delta^a\,
    e^{-S_{eff}[\Delta^*,\Delta]},
\end{equation}
with the effective action
\begin{equation}
\label{Seff a}
    S_{eff}=-\ln\langle\Det{\cal G}^{-1}\rangle+
    \frac{1}{g}\sum_a\int\limits_0^\beta d\tau
    \sum\limits_{\br\br'}|\Delta^a_{\br\br'}|^2,
\end{equation}
instead of Eq. (\ref{Seff gen}). Here
\begin{equation}
\label{Gab}
    {\cal G}_{ab}^{-1}(\br,\tau;\br',\tau')
    =\delta_{ab}\delta(\tau-\tau')\left(\begin{array}{cc}
    \delta_{\br\br'}(-\partial_{\tau}-U_{\br})-\xi_{\br\br'} &
    -\Delta^a_{\br\br'}(\tau)\\
    -\Delta^{a,*}_{\br\br'}(\tau) &
    \delta_{\br\br'}(-\partial_{\tau}+U_{\br})
    +\xi_{\br\br'}
    \end{array}\right),
\end{equation}
and ``Det'' stands for the full operator determinant with respect
to the space-time coordinates and the Nambu and the replica
indices. The disorder averaging in the first term in $S_{eff}$
generates effective coupling between different replicas, resulting
in a non-linear bosonic field theory.

While the saddle point of the effective action (\ref{Seff a}) has
the same structure as in the clean case:
$\Delta^a_{0,\br\br'}(\tau)=\Delta_{0,\br\br'}$, see Eq.
(\ref{Delta dwave}), the temperature dependence of the gap
amplitude is different, in particular, the critical temperature is
suppressed by impurities.\cite{Book} The mean-field Green's
function is the unity matrix in the replica space: ${\cal
G}_{0,ab}^{-1}=\delta_{ab}{\cal G}_0^{-1}$, where
\begin{equation}
\label{cal G0 dis}
    {\cal G}_0^{-1}(\br,\br';\omega_n)=
    \left(\begin{array}{cc}
    \delta_{\br\br'}(i\omega_n-U_{\br})-\xi_{\br\br'} &
    -\Delta_{0,\br\br'}\\
    -\Delta_{0,\br\br'} &
    \delta_{\br\br'}(i\omega_n+U_{\br})
    +\xi_{\br\br'}\end{array}\right)
\end{equation}
in the Matsubara frequency representation. The disorder averaging
can be done using the standard diagram technique.\cite{AGD} In the
so-called self-consistent Born approximation, which assumes a
sufficiently weak impurity scattering and also neglects the
diagrams with crossed impurity lines, the average mean-field
Green's function has the form
\begin{equation}
\label{G0 dis}
    \langle{\cal G}_0(\bk,\omega_n)\rangle=
    -\frac{i\tilde\omega_n\tau_0+\xi_{\bk}\tau_3+\Delta_{\bk}\tau_1}{\tilde\omega_n^2
    +\xi_{\bk}^2+\Delta_{\bk}^2},
\end{equation}
where $\tilde\omega_n$ satisfies the equation
$i\tilde\omega_n=i\omega_n+(1/2\tau)
\langle(i\tilde\omega_n/\sqrt{\tilde\omega_n^2+\Delta_{\bk}^2})\rangle_{\bk}$.\cite{Book}

Assuming replica-symmetric phase fluctuations, the order parameter
can be written in the form
$\Delta^a_{\br\br'}(\tau)=\Delta_{0,\br\br'}e^{i[\theta_{\br}(\tau)+\theta_{\br'}(\tau)]/2}$,
see Eq. (\ref{average phase}). Performing the gauge transformation
(\ref{tilde G}) in each replica, we have $\tilde{\cal
G}_{ab}^{-1}=\delta_{ab}({\cal G}_0^{-1}-\Sigma)$, where the
self-energy operator has the form (\ref{Sigma}) with
$\varphi_{\br}(\tau)=0$. Therefore,
$$
    \ln\langle\Det{\cal G}^{-1}\rangle=
    \ln\langle\Det{\tilde{\cal G}}^{-1}\rangle
    =n\langle\Tr\ln{\cal G}_0^{-1}\rangle+
    n\langle\Tr\ln(1-{\cal G}_0\Sigma)\rangle+O(n^2),
$$
in the limit $n\to 0$. Substituting this expansion in Eq.
(\ref{Zn}), we see that the replica index can be omitted, and the
effective action (\ref{Seff theta phi}) gets replaced by its
disorder average:
\begin{equation}
\label{Seff phase dis}
    S_{eff}[\theta]=\beta{\cal E}_0-\langle\Tr\ln(1-{\cal G}_0\Sigma)\rangle,
\end{equation}
where ${\cal E}_0$ is the mean-field energy of a disordered
superconductor. Considering only the static fluctuations and
expanding the operator trace in powers of the phase gradients, we
obtain the effective action in the form (\ref{Seff cl}), where the
energy of fluctuations is now given by
\begin{equation}
\label{cal E dis}
    {\cal E}=\frac{1}{2}
    \sum\limits_{\bq}\;\langle{\cal K}_{ij}(\bq)\rangle
    v^*_{s,i}(\bq)v_{s,j}(\bq),
\end{equation}
with
\begin{equation}
\label{kernel av}
    \langle{\cal K}_{ij}(\bq)\rangle=
    \langle{\cal K}^{(0)}_{ij}\rangle+\langle{\cal I}_{ij}(\bq)\rangle.
\end{equation}
Before proceeding with the calculation of the disorder averages,
we would like to note that Eq. (\ref{cal E dis}) could also be
derived using a less formal approach, without introducing
replicas. Assuming that electrons move in the presence of the
random potential and a given order parameter field
$\Delta_{\br\br'}=\Delta_{0,\br\br'}e^{i(\theta_{\br}+\theta_{\br'})/2}$,
one can define the energy functional of phase fluctuations:
\begin{equation}
\label{GL func}
    {\cal E}[\theta]=-\frac{1}{\beta}\langle\ln Z[\theta]\rangle+
    \frac{1}{\beta}\langle\ln Z[0]\rangle,
\end{equation}
where $Z[\theta]=\Det({\cal G}_0^{-1}-\Sigma)$ is the partition
function. Expanding $\ln Z[\theta]$ in powers of the phase
gradients, followed by averaging each expansion coefficient with
respect to disorder, we arrive at Eq. (\ref{cal E dis}).

The disorder averaging of the first term in Eq. (\ref{kernel av})
is straightforward:
\begin{equation}
\label{cal K0 dis}
    \langle{\cal K}^{(0)}_{ij}\rangle=
    2m^2T\sum\limits_n\int\frac{d^D\bk}{(2\pi)^D}\,
    m^{-1}_{ij}(\bk)\langle G_0(\bk,\omega_n)\rangle=
    \frac{m}{m^*}\rho_0\delta_{ij},
\end{equation}
where we used the effective mass approximation, as in Eq.
(\ref{cal K0 eff mass}).

The average of the product of two Green's functions in the second
term includes the impurity vertex corrections. However, since we
are interested in the behavior of the kernel in the
long-wavelength limit $\bq\to 0$ in a sufficiently clean system
(the precise criterion will be discussed below), the vertex
corrections are negligible. Thus
\begin{equation}
\label{cal I dis}
    \langle{\cal I}_{ij}(\bq)\rangle=m^2
    T\sum\limits_n\int\frac{d^D\bk}{(2\pi)^D}v_i(\bk)
    v_j(\bk)\,\tr[\langle{\cal
    G}_0(\bk+\bq,\omega_n)\rangle\tau_0
    \langle{\cal G}_0(\bk,\omega_n)\rangle\tau_0].
\end{equation}
In order to calculate the Matsubara sum we use the spectral
representation
\begin{equation}
\label{spec rep}
    {\cal G}_0(\bk,\omega_n)=\int\limits_{-\infty}^{\infty}\frac{d\epsilon}{\pi}
    \frac{\im{\cal G}_0^R(\bk,\epsilon)}{\epsilon-i\omega_n},
\end{equation}
where the retarded matrix Green's function is obtained from Eq.
(\ref{G0 dis}) by analytical continuation:
\begin{equation}
\label{G R}
    {\cal G}_0^R(\bk,\epsilon)=\frac{t(\epsilon)\tau_0+\xi_{\bk}\tau_3+
    \Delta_{\bk}\tau_1}{t^2(\epsilon)-\xi_{\bk}^2-\Delta_{\bk}^2},
\end{equation}
and $t(\epsilon)=(i\tilde\omega_n)|_{i\omega_n\to\epsilon+i0}$
satisfies the equation
$t=\epsilon+(i/2\tau)\langle(t/\sqrt{t^2-\Delta_{\bk}^2})\rangle_{\bk}$.
Instead of finding the exact energy dependence of $t(\epsilon)$
for the $d$-wave gap, we use below a simplified expression
\begin{equation}
\label{t}
    t(\epsilon)=\epsilon+i\Gamma,
\end{equation}
which captures the essential qualitative effects of the impurity
broadening of the single-electron states, with $\Gamma$ being an
energy-independent effective scattering rate. The exact solution
shows that the scattering rate indeed approaches a constant value
at $\epsilon\to 0$, i.e. in the so-called universal
limit.\cite{Book}

Inserting the representation (\ref{spec rep}) in Eq. (\ref{cal I
dis}), one obtains
\begin{eqnarray}
    \langle{\cal I}_{ij}(\bq)\rangle=-\frac{m^2}{2}
    \int\limits_{-\infty}^\infty d\epsilon_1\int\limits_{-\infty}^\infty d\epsilon_2
    \frac{\tanh\frac{\epsilon_1}{2T}-\tanh\frac{\epsilon_2}{2T}}{\epsilon_1-\epsilon_2}
    \int\frac{d^D\bk}{(2\pi)^D}\,v_i(\bk)v_j(\bk)\nonumber\\
    \times\bigl\{C_-(\bk,\bq)[d_+(\bk+\bq,\epsilon_1)d_-(\bk,\epsilon_2)+
    d_-(\bk+\bq,\epsilon_1)d_+(\bk,\epsilon_2)]\nonumber\\
    \label{cal I dis gen} +C_+(\bk,\bq)[d_+(\bk+\bq,\epsilon_1)d_+(\bk,\epsilon_2)+
    d_-(\bk+\bq,\epsilon_1)d_-(\bk,\epsilon_2)]\bigr\},
\end{eqnarray}
where
$$
    d_\pm(\bk,\epsilon)=\frac{1}{\pi}\frac{\Gamma}{
    (\epsilon\pm E_{\bk})^2+\Gamma^2}.
$$
and $C_\pm$ are the coherence factors (\ref{Cpm}). The expressions
(\ref{cal I}) are recovered from Eq. (\ref{cal I dis gen}) in the
clean limit $\Gamma\to 0$, when
$d_\pm(\bk,\epsilon)\to\delta(\epsilon\pm E_{\bk})$.

We focus on the case of zero temperature, when the kernel is a
non-analytical function of $\bq$ in the absence of impurities, see
Eq. (\ref{K zero T}). At $T=0$, the integrals over
$\epsilon_{1,2}$ can be calculated, giving
\begin{equation}
    \langle{\cal I}_{ij}(\bq)\rangle=
    -\frac{2m^2}{\pi}\int\frac{d^D\bk}{(2\pi)^D}\,v_i(\bk)v_j(\bk)
    \biggl[C_-(\bk,\bq)\frac{\arctan\frac{E_{\bk+\bq}}{\Gamma}+
    \arctan\frac{E_{\bk}}{\Gamma}}{E_{\bk+\bq}+E_{\bk}}
    +C_+(\bk,\bq)\frac{\arctan\frac{E_{\bk+\bq}}{\Gamma}-
    \arctan\frac{E_{\bk}}{\Gamma}}{E_{\bk+\bq}-E_{\bk}}\biggr].
\end{equation}
Comparing this to Eq. (\ref{cal I}), we see that the energy of
classical phase fluctuations in the disordered case at zero
temperature has exactly the same form as in the clean case at a
finite temperature, if one formally replaces
$\tanh(E/2T)\to(2/\pi)\arctan(E/\Gamma)$. Therefore one can expect
that the disorder will affect the phase fluctuations in the same
way as temperature does, i.e. the singularities of the effective
action will be washed out.

To check this conclusion quantitatively, let us evaluate the
long-wavelength asymptotics of $\langle{\cal I}_{ij}(\bq)\rangle$
in the nodal approximation. Using Eqs.
(\ref{nodal_approx},\ref{nodal int}), we have
\begin{equation}
\label{cal I dis sum}
    \langle{\cal I}_{ij}(\bq)\rangle=-\frac{m^2}{2\pi}\frac{v_F}{v_\Delta}
    \sum\limits_{n=1}^4\hat k_{n,i}\hat k_{n,j}\tilde S_n(\bq),
\end{equation}
where
$$
    \tilde S_1(\bq)=\tilde S_3(\bq)=\Gamma\tilde s\left(\frac{\gamma_{1}}{\Gamma}\right),\quad
    \tilde S_2(\bq)=\tilde S_4(\bq)=\Gamma\tilde
    s\left(\frac{\gamma_{2}}{\Gamma}\right),
$$
and $\gamma_{1,2}(\bq)$ are given by Eq. (\ref{gamma12}). The
function $\tilde s(x)$ is defined by
\begin{equation}
\label{s dis}
    \tilde s(x)=\int\limits_0^{2\pi}\frac{d\alpha}{2\pi}
    \int\limits_0^{y_{max}} dy\,y[\tilde f_+(x,y,\alpha)+\tilde f_-(x,y,\alpha)],
\end{equation}
where $y_{max}=\epsilon_{max}/\Gamma$, and
$$
   \tilde f_\pm=\frac{1}{\pi}\left(1\pm\frac{y+
   x\cos\alpha}{\sqrt{x^2+y^2+2xy\cos\alpha}}\right)
   \frac{\arctan(\sqrt{x^2+y^2+2xy\cos\alpha})\mp
   \arctan(y)}{\sqrt{x^2+y^2+2xy\cos\alpha}\mp y}.
$$
In contrast to the clean case, see Eq. (\ref{s}), the energy
cutoff here cannot be extended to infinity, due to the logarithmic
divergence of the integrals.

The clean-case expression (\ref{K zero T}) is recovered from Eq.
(\ref{s dis}) at $x\gg 1$, i.e. at
$\Gamma\ll\gamma_{1,2}(\bq)\ll\epsilon_{max}$. In the opposite
limit $x\ll 1$, one finds
\begin{equation}
\label{tilde s asymp}
    \tilde s(x)=\frac{2}{\pi}\ln y_{max}+\frac{1}{6\pi}x^2,
\end{equation}
which is valid at $y_{max}\gg 1$. Inserting this into Eq.
(\ref{cal I dis sum}), we finally obtain
\begin{equation}
\label{K Gamma}
    \langle{\cal K}_{ij}(\bq)\rangle=\left(\frac{m}{m^*}\rho_0
    -\frac{2}{\pi^2}\frac{v_F}{v_\Delta}m^2
    \Gamma\ln\frac{\epsilon_{max}}{\Gamma}\right)\delta_{ij}
    -\frac{m^2}{12\pi^2}\frac{v_F}{v_\Delta}\frac{1}{\Gamma}
    \left(\begin{array}{cc}
    (v_F^2+v_\Delta^2)(q_x^2+q_y^2) & 2(v_F^2-v_\Delta^2)q_xq_y\\
    2(v_F^2-v_\Delta^2)q_xq_y & (v_F^2+v_\Delta^2)(q_x^2+q_y^2)
    \end{array}\right)_{ij}.
\end{equation}
The first term describes the depletion of the superfluid density
at zero temperature due to the impurity scattering:
\begin{equation}
\label{sf density Gamma}
    \rho_s(T=0,\Gamma)=\rho_s(T=0,0)-\frac{2}{\pi^2}\frac{v_F}{v_\Delta}m^2
    \Gamma\ln\frac{\Delta_0}{\Gamma},
\end{equation}
see also Ref. \cite{HG93} (here we used
$\epsilon_{max}\simeq\Delta_0$). Upon increasing $\Gamma$, the
superfluid density decreases and eventually vanishes at some
critical disorder strength. One cannot reach the critical point
using the expression (\ref{sf density Gamma}), since its validity
is limited to the case of weak impurity scattering
$\Gamma\ll\Delta_0$.

The second term in Eq. (\ref{K Gamma}) shows that, in contrast to
the clean case, the energy is a non-singular function of momentum,
even at $T=0$. Comparing it to the expression (\ref{XY kernel}),
we see that the microscopic theory, at least at the Gaussian
level, has the same long-wavelength behavior as the classical
$XY$-model. The non-zero elements of the tensor $\Lambda$ are
given by
\begin{equation}
\label{Lambda micro}
    \begin{array}{l}
    \displaystyle\Lambda_{xx,xx}=\Lambda_{xx,yy}=-
    \frac{m^2}{12\pi^2}\frac{v_F}{v_\Delta}\frac{1}{\Gamma}(v_F^2+v_\Delta^2),\\
    \displaystyle \Lambda_{xy,xy}=
    -\frac{m^2}{12\pi^2}\frac{v_F}{v_\Delta}\frac{1}{\Gamma}(v_F^2-v_\Delta^2), \\
    \end{array}
\end{equation}
with other elements obtained by symmetry. One can see that keeping
only the nearest-neighbor phase stiffness coefficient in Eq.
(\ref{Lambda}) is not sufficient to reproduce the tensor structure
(\ref{Lambda micro}), therefore one has to consider an extended
$XY$-model, in which the couplings between next-nearest neighbors
\emph{etc}, are also taken into account. This was first noticed in
Ref. \cite{KC02} for an $s$-wave superconductor.

\section{Conclusions}
\label{sec:Conclusion}

To summarize, we have shown that the effective field theory for
classical phase fluctuations in a clean $d$-wave superconductor
suffers from singularities which make the gradient expansion of
the fluctuation energy impossible. This means, in particular, that
the physics of classical phase fluctuations cannot be described by
the $XY$-model at low temperatures. In the presence of disorder, a
well-defined gradient expansion of the Gaussian phase-only action
is restored, and has the same form as that of an extended
classical $XY$-model.

\section*{Acknowledgements}

We are grateful to S. Sharapov for useful discussions. This work
was supported by the Natural Sciences and Engineering Research
Council of Canada.

\appendix

\section{Condensate energy and non-linear Meissner effect}
\label{app:Int}

In this Appendix we calculate the higher-order terms in the
expansion of the effective action
\begin{equation}
\label{Seff diff}
    S_{eff}=S_0-\Tr\ln\tilde{\cal G}^{-1}+\Tr\ln {\cal G}_0^{-1},
\end{equation}
in powers of the phase gradients. We consider only the limit of
uniformly-moving condensate, when the superfluid velocity $\bvs$
is constant, so that
$\theta_{\br}-\theta_{\br'}=2m\bvs(\br-\br')$. In the absence of
the scalar potential, the self-energy (\ref{Sigma}) becomes
translationally invariant:
$$
    \Sigma(\br,\tau;\br',\tau')
    =\delta(\tau-\tau')\tau_3\xi_{\br\br'}
    \bigl[e^{-i\tau_3m\bvs(\br-\br')}-1\bigr].
$$
The gauge-transformed Green's function (\ref{tilde G}) becomes
diagonal in the momentum space:
\begin{equation}
    \tilde{\cal G}^{-1}(\bk,\omega_n)=\left(\begin{array}{cc}
    i\omega_n-\xi^+_{\bk} & -\Delta_{\bk} \\
    -\Delta_{\bk} & i\omega_n+\xi^-_{\bk}
    \end{array}\right),
\end{equation}
where $\xi^{\pm}_{\bk}=\xi_{\bk\pm m\bvs}$, which allows one to
calculate the operator traces:
$$
    \Tr\ln\tilde{\cal G}^{-1}=\ln\Det\tilde{\cal G}^{-1}
    =\beta{\cal V} T\sum\limits_n\int\frac{d^D\bk}{(2\pi)^D}\,
    \ln\det\tilde {\cal G}^{-1}(\bk,\omega_n)
$$
where ${\cal V}$ is the system volume and ``det'' denotes a
$2\times 2$ matrix determinant in the electron-hole space.
Inserting this in Eq. (\ref{Seff diff}), we obtain
$S_{eff}=S_0+\beta{\cal E}$, where
\begin{equation}
\label{Seff nl}
    {\cal E}=-{\cal V}T\sum\limits_n\int\frac{d^D\bk}{(2\pi)^D}
    \ln\frac{(i\omega_n-\tilde E_{\bk,+})
    (i\omega_n+\tilde E_{\bk,-})}{(i\omega_n-E_{\bk})(i\omega_n+E_{\bk})}
\end{equation}
has the meaning of the kinetic energy of uniformly moving
condensate, and
\begin{equation}
    \tilde E_{\bk,\pm}=\pm\frac{\xi^+_{\bk}-\xi^-_{\bk}}{2}+
    \sqrt{\left(\frac{\xi^+_{\bk}+\xi^-_{\bk}}{2}\right)^2+\Delta_{\bk}^2}
\end{equation}
are the quasiparticle energies affected by the superflow. Using
the identity
$$
    T\sum\limits_n\ln\frac{i\omega_n-\tilde E}{i\omega_n-E}
    =T\ln\frac{\cosh\frac{\tilde E}{2T}}{\cosh\frac{E}{2T}},
$$
the energy density can be written in the form
\begin{equation}
\label{E int}
    \frac{\cal E}{\cal V}=-T\int\frac{d^D\bk}{(2\pi)^D}\;
    \ln\frac{\cosh\frac{\tilde E_{\bk,+}}{2T}
    \cosh\frac{\tilde E_{\bk,-}}{2T}}{\cosh^2\frac{E_{\bk}}{2T}}.
\end{equation}
At $T=0$, this becomes
\begin{equation}
\label{E zeroT}
    \frac{\cal E}{\cal
    V}=-\frac{1}{2}\int\frac{d^D\bk}{(2\pi)^D}\,
    \bigl(|\tilde E_{\bk,+}|+|\tilde E_{\bk,+}|-2E_{\bk}\bigr).
\end{equation}

We focus now on the case of a two-dimensional $d$-wave
superconductor. Assuming for simplicity a Galilean-invariant case,
characterized by the parabolic dispersion
$\xi_{\bk}=\bk^2/2m-\mu$, with the effective mass equal to the
bare electron mass $m$, we have
$$
    \tilde E_{\bk,\pm}=\sqrt{(\xi_{\bk}+\zeta)^2+
    \Delta_{\bk}^2}\pm\bk\bvs=E_{\zeta,\bk}\pm\bk\bvs,
$$
where $\bk\bvs$ is the so-called Doppler shift of the
quasiparticle energy in the presence of moving condensate, and
$\zeta=m\bvs^2/2$. The expression (\ref{E zeroT}) can then be
written as ${\cal E}/{\cal V}=({\cal E}/{\cal V})_{reg}+({\cal
E}/{\cal V})_{Dopp}$. The first term,
\begin{equation}
\label{E reg}
    \left(\frac{\cal E}{\cal V}\right)_{reg}=-\int\frac{d^2\bk}{(2\pi)^2}\;
    (E_{\zeta,\bk}-E_{\bk})
    =-\zeta\int\frac{d^2\bk}{(2\pi)^2}\;\frac{\xi_{\bk}}{E_{\bk}}+
    O(\zeta^2)=\frac{\rho_0\bvs^2}{2}+O(\bvs^4),
\end{equation}
has a well-defined Taylor expansion in powers of $\bvs$. Note that
the nodal approximation cannot be used here because of the
contributions from the regions far from the Fermi surface, see
Sec. \ref{sec:2D-class}. In contrast, the second term,
\begin{equation}
\label{E Dopp}
    \left(\frac{\cal E}{\cal V}\right)_{Dopp}=-\frac{1}{2}
    \int\frac{d^2\bk}{(2\pi)^2}\;\Bigl[|E_{\zeta,\bk}+\eta|+
    |E_{\zeta,\bk}-\eta|-2E_{\zeta,\bk}\Bigr],
\end{equation}
can be calculated in the nodal approximation. Using Eqs.
(\ref{nodal_approx},\ref{nodal int}), we obtain
$$
    \left(\frac{\cal E}{\cal V}\right)_{Dopp}=-\frac{1}{12\pi
    v_Fv_\Delta}\sum\limits_{n=1}^4|\bk_n\bvs|^3,
$$
which is a non-analytical function of $\bvs$. Putting all pieces
together, we arrive at the following result:
\begin{equation}
\label{E result}
    \frac{\cal E}{\cal V}=\frac{\rho_0\bvs^2}{2}-
    \frac{m^3v_F^2}{12\sqrt{2}\pi v_\Delta}\bigl(|v_{s,x}+v_{s,y}|^3+
    |v_{s,x}-v_{s,y}|^3\bigr)+O(\bvs^4).
\end{equation}
We see that the nodal quasiparticles make the kinetic energy of
the condensate a non-analytical function of the superfluid
velocity.\cite{Vol97} This singular behavior is closely related to
the so-called non-linear Meissner effect:\cite{non-lin} the
screening supercurrent acts as a pair-breaker in $d$-wave
superconductors, creating a finite density of normal excitations
even at $T=0$. The depletion of the supercurrent by those
excitations leads to a non-analytical dependence of the
electromagnetic response functions on $\bvs$ and therefore the
external magnetic field.




\begin{thebibliography}{99}

\bibitem{LarVar05}
A. I. Larkin and A. A. Varlamov, \emph{Theory of Fluctuations in
Superconductors} (Oxford University Press, 2005).

\bibitem{KBCC88}
A. Kapitulnik, M. R. Beasley, C. Castellani, and C. Di Castro,
Phys. Rev. B \textbf{37}, 537 (1988).

\bibitem{TS99}
T. Timusk and B. Statt, Rep. Progr. Phys.
\textbf{62}, 61 (1999).

\bibitem{EK95-2}
V. J. Emery and S. A. Kivelson, Nature (London)
\textbf{374}, 434 (1995).

\bibitem{BKT}
V. L. Berezinskii, Zh. Eksp. Teor. Fiz. \textbf{61}, 1144 (1971)
[Sov. Phys. -- JETP \textbf{34}, 610 (1972)]; J. M. Kosterlitz and
D. J. Thouless, J. Phys. C \textbf{6}, 1181 (1973).

\bibitem{LQS01}
V. M. Loktev, R. Quick, and S. G. Sharapov, Phys.
Rep. \textbf{349}, 1 (2001).

\bibitem{EK95-1}
V. J. Emery and S. A. Kivelson, Phys. Rev. Lett.
\textbf{74}, 3253 (1995).

\bibitem{CKEM99}
E. W. Carlson, S. A. Kivelson, V. J. Emery, and
E. Manousakis, Phys. Rev. Lett. \textbf{83} 612 (1999).

\bibitem{KDH01}
H.-J. Kwon, A. T. Dorsey, and P. J. Hirschfeld,
Phys. Rev. Lett. \textbf{86}, 3875 (2001).

\bibitem{BCCPR01}
L. Benfatto, S. Caprara, C. Castellani, A.
Paramekanti, and M. Randeria, Phys. Rev. B \textbf{63}, 174513
(2001).

\bibitem{SM03}
K. V. Samokhin and B. Mitrovi\'c, Phys. Rev. Lett.
\textbf{92}, 057002 (2004).

\bibitem{Book}
V. P. Mineev and K. V. Samokhin, \emph{Introduction
to Unconventional Superconductivity} (Gordon and Breach, 1999).

\bibitem{AGD}
A. A. Abrikosov, L. P. Gorkov, and I. E.
Dzyaloshinski, \emph{Methods of Quantum Field Theory in
Statistical Physics} (Dover Publications, 1975).

\bibitem{BTCC02}
L. Benfatto, A. Toschi, S. Caprara, and C.
Castellani, Phys. Rev. B \textbf{66}, 054515 (2002).

\bibitem{PRRM00}
A. Paramekanti, M. Randeria, T. V. Ramakrishnan,
and S. S. Mandal, Phys. Rev. B \textbf{62}, 6786 (2000).

\bibitem{Lee93}
P. A. Lee, Phys. Rev. Lett. \textbf{71}, 1887
(1993).

\bibitem{ratio}
M. Chiao, R. W. Hill, C. Lupien, L. Taillefer, P.
Lambert, R. Gagnon, and P. Fournier, Phys. Rev. B \textbf{62},
3554 (2000).

\bibitem{DL00}
A. Durst and P. A. Lee, Phys. Rev. B \textbf{62},
1270 (2000).

\bibitem{LW97}
P. A. Lee and X.-G. Wen, Phys. Rev. Lett.
\textbf{78}, 4111 (1997).

\bibitem{Gross86}
F. Gross, B. S. Chandrasekhar, D. Einzel, K.
Andres, P. J. Hirschfeld, H. R. Ott, J. Beuers, Z. Fisk, and J. L.
Smith, Z. Phys. B \textbf{64}, 175 (1986).

\bibitem{AGR91}
J. Annett, N. Goldenfeld, and S. R. Renn, Phys.
Rev. B \textbf{43}, 2778 (1991).

\bibitem{Hardy93}
W. N. Hardy, D. A. Bonn, D. C. Morgan, R. Liang,
and K. Zhang, Phys. Rev. Lett. \textbf{70}, 3999 (1993).

\bibitem{Tink96}
M. Tinkham, {\em Introduction to Superconductivity} (McGraw-Hill,
New York, 1996).

\bibitem{non-loc}
I. Kosztin and A. J. Leggett, Phys. Rev. Lett. {\bf
79}, 135 (1997); M. Franz, I. Affleck, and M. H. S. Amin, Phys.
Rev. Lett. {\bf 79}, 1555 (1997).

\bibitem{GW84}
\emph{Percolation, localization, and superconductivity}, A. M.
Goldman and S. A. Wolf (eds.) (Plenum Press, 1984).


\bibitem{BTC04}
L. Benfatto, A. Toschi, and S. Caprara,
Phys. Rev. B \textbf{69}, 184510 (2004) .

\bibitem{sigma-models}
I. V. Lerner, preprint cond-mat//0307471.

\bibitem{BK94}
D. Belitz and T. R. Kirkpatrick, Rev. Mod. Phys.
\textbf{66}, 261 (1994).

\bibitem{HG93}
P. J. Hirschfeld and N. Goldenfeld, Phys. Rev. B \textbf{48},
4219(R) (1993).

\bibitem{KC02}
W. Kim and J. P. Carbotte, Europhys. Lett.
\textbf{59}, 761 (2002).

\bibitem{Vol97}
G. E. Volovik, Pis'ma Zh. Eksp. Teor. Fiz. \textbf{65}, 465 (1997)
[JETP Lett. \textbf{65}, 491 (1997)].

\bibitem{non-lin}
S. K. Yip and J. A. Sauls, Phys. Rev. Lett. {\bf 69}, 2264 (1992);
D. Xu, S. K. Yip, and J. A. Sauls, Phys. Rev. B {\bf 51}, 16233
(1995).



\end{thebibliography}
\end{document}